\documentclass[prl,showpacs,footinbib,twocolumn,final]{revtex4}
\usepackage{graphics,graphicx,color,amstext,amsxtra,amssymb,latexsym,epsf,showkeys}

\newcommand{\be}{\begin{equation}}
\newcommand{\ee}{\end{equation}}
\newcommand{\bs }{\boldsymbol }

\renewcommand{\r}{{\boldsymbol r}}

\newcommand{\limite}[1]{ {{\raisebox{-.3cm}{$\textstyle\longrightarrow$}} \atop {\scriptstyle{#1}}}}
\definecolor{gris}{gray}{0.9}

\begin{document}
\title{Interference pattern of a long diffusive Josephson junction}

\author{Gilles Montambaux}
\affiliation{Laboratoire de Physique des Solides, UMR8502, CNRS, Universit\'{e}
Paris-Sud, CNRS,  91405 Orsay Cedex, France}

\date{July 3, 2007}


\begin{abstract}
We calculate the modulation by a magnetic field of the critical
current  of a long disordered Josephson junction in the diffusive
limit, i.e. when the dimensions of the junction are larger that
the elastic mean free path, and when the length $L$ is much larger
than the width $w$. Due to the averaging of the gauge invariant
phase factor over diffusive trajectories, the well-known
oscillations of the Fraunhofer pattern are smoothed out and
replaced by  an exponential decay at large field. The predicted
pattern is universal, i.e.,  it is independent of the disorder
strength. We point out an interesting relation with the physics of
speckle correlations in optics of turbid media.
\end{abstract}

\maketitle

{\it Introduction - } The supercurrent flowing through a tunnel
junction between two superconductors is given by the well-known
gauge invariant Josephson relation
\begin{equation}
I(\delta)=I_0\sin \left(\delta - {2 e \over \hbar} \int {\bs A}
\cdot d {\bs l}\right) \label{JJ} \end{equation} 
where $\delta$ is
the phase difference between the two superconductors. The
Josephson effect is thus a beautiful tool to exhibit interference
effects, manifestations of the phase coherence of the
superconducting wavefunction. For example, a circuit with two
Josephson junctions is a realization of Young's two slits
experiment, where the interference is modulated by the
Aharonov-Bohm flux through the circuit \cite{Mercereau64}.
Moreover a single Josephson junction with a finite width exhibits
an interference pattern reminiscent of the diffraction
(Fraunhofer) pattern of a slit, as recalled in eq. (\ref{sinc})
\cite{Mercereau65}.

\begin{figure}[h]
\includegraphics[scale=1.2]{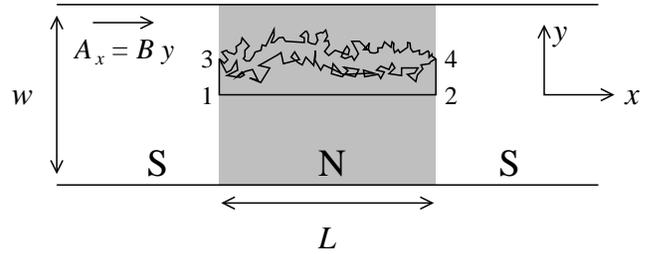}
  \caption{Geometry discussed in the
text. The junction (grey) is a diffusive metal so that the current
through the junction results form the contribution of many
diffusive trajectories. We assume that the junction has a quasi-1D
geometry: $L \gg w$.}\label{jonction}
\end{figure}

It is natural to wonder whether such an interference experiment
can probe phase coherence in a more complex medium with multiple
scattering of the electrons. Here, we consider a long Josephson
junction made of a diffusive metal forming a quasi-one-dimensional
wire. The junction of length $L$ is attached to superconducting
leads along the direction $x$. A magnetic field is applied along
the direction $z$ perpendicular to the wire. The width of the
junction (along the $y$ direction) is denoted by $w$, and its
width (along $z$) is denoted by $h$. We consider a long junction
such that $L \gg w$. The junction is schematically represented on
figure \ref{jonction}. The amplitude of the Josephson current in
such a diffusive junction has been calculated with the Usadel
equation \cite{Belzig96,Heikkila02}. It has been found that,
contrary to the the case of the tunnel junction, for a good
contact between the metallic region and the superconductors, the
Josephson relation may not be sinusoidal. However, harmonics are
expected to decay rapidly, roughly as $(-1)^n /n^2$. In this
letter, we assume a sinusoidal Josephson relation and consider how
the phase is modified by the application of a magnetic field. We
find that for a long diffusive junction, the critical current
varies as
\begin{equation} I_c = I_0 { {\pi  \over \sqrt{3}} {\phi \over \phi_0 }  \over
\sinh  {\pi  \over \sqrt{3}} {\phi \over \phi_0 } }\label{r2}
\end{equation}
where $\phi$ is the flux through the junction and $\phi_0=h/2e$ is
the superconducting flux quantum.
\medskip

{\it Josephson current and diffusive trajectories - }
Quite generally, the Josephson current resulting from all current
paths has the form

$$I(\delta)= I_0 \left\langle \sin \left(\delta({\cal C}) - {2 e \over \hbar} \int_{\cal C} {\bs A} \cdot d
{\bs l}\right) \right\rangle_{\cal C}$$ where $\langle \cdots
\rangle_{\cal C}$ denotes the average over all current paths
through the junction.

We choose a gauge where the vector potential ${\bs A}$  is aligned
along the direction $x$, $A_x=By$, $y \in [-w/2,w/2]$. On figure
\ref{jonction}, the diffuse paths  $3-4$ represent  current paths
${\cal C}$ while the straight path $1-2$ serves as a reference
path and corresponds to $y=0$. The circulation of ${\bs A}$ is
zero along the paths $1-2$, $1-3$ and $2-4$, so that the phase
difference does not depend on $y$ and is denoted $\delta_0$. The
current can be rewritten in the gauge dependent form

$$I(\delta)= I_0 \left\langle \sin \left(\delta_0 - {2 e \over \hbar} \int_{\cal C} {\bs A} \cdot d
{\bs l}\right) \right\rangle_{\cal C}$$ so that we write the
critical current $I_c=\max[I(\delta)]$ as
\begin{equation} I_c=I_0 \left|\left\langle \displaystyle e^{-i{2 \pi \over \phi_0} \int_{\cal C}
\bs{A} d {\bs l}} \right\rangle_{\cal C}\right| \label{Icgene}
\end{equation}
or, in a gauge independent form:
$$I_c=I_0 \left|\left\langle e^{-i{2 \pi \phi({\cal C}) \over \phi_0} } \right\rangle_{\cal
C}\right| \  . $$
$\phi({\cal C})$ is the flux through the area $1-3-4-2$ defined by
a diffusive path ${\cal C}$. $\phi_0=h/2 e$ is the superconducting
flux quantum. We have neglected the current flowing in the
superconductors along the $1-3$ and $2-4$ segments, {\it i.e.} we
have assumed that the penetration length $\lambda$ of the magnetic
field in the superconductor $\lambda \rightarrow 0$. Taking into
account a finite $\lambda$ would amount to replace $L$ by $L+2
\lambda$, as usual.
\medskip

As a reminder, we first briefly consider the case of the short
ballistic junction. The current has to be summed on the current
paths
$$I_c=I_0 \left\langle e^{-{{2 i \pi \over \phi_0 } B y L}}\right\rangle={I_0 \over w} \int_{-w/2}^{w/2} e^{-{{2 i \pi
\over \phi_0 } B y L}} dy$$ leading to the well-known
Fraunhofer-like result \cite{Mercereau65}
\begin{equation} I_c=I_0 \left|{\sin \pi \phi /\phi_0 \over \pi
\phi/\phi_0}\right| \label{sinc}
\end{equation} 
where $\phi=B w L$ (or $\phi=w(L+2 \lambda)$ to account for a finite penetration length).

For a long diffusive junction, the phase factor in (\ref{Icgene})
has to be averaged on the distribution of diffusive trajectories.
In order to perform this average,   we need to describe the
diffusion from a point $\r$ at one end of the diffusive sample to
another point $\r'$ located at the other end. We introduce the
probability $P(\r,\r',t)$, solution of the covariant equation

\begin{equation}
\left[ {\partial \over \partial t} - D\left( \nabla_{\r'} + i {2 e
\over \hbar} {\bs A}(\r') \right)^2\right] P(\r,\r',t) =
\delta(\r-\r') \delta(t) \label{eqdiffcoop}\end{equation}
where the electron charge is denoted $-e$. This solution may be
expressed as a functional integral \cite{Feynman}~:
\begin{equation}
P(\r, \r', t) =  \int_{\r (0) = \r}^{\r (t) = \r'} \hspace{-.5cm}
{\cal D}\{ \r \}  \exp \left( - \! \int_0^t \left[{{\dot \r} ^2
\over 4 D} + i{ 2 e \over \hbar} {\dot  \r} . {\bs A}(\r) \right]
d \tau \right) \label{eqdifffonct}
\end{equation}

 We consider a long junction where the dephasing between $\r$ and
$\r'$ is supposed to be independent of the position of $\r$ and
$\r'$ on the boundaries, that is independent of the coordinates
$y$ and $z$. Therefore, we consider a one-dimensional diffusion
equation with the appropriate gauge:
\begin{equation}
\left[ {\partial \over \partial t} - D\left( {\partial \over
\partial x} + i {2 e \over \hbar} B y \right)^2\right]
P(x,x',t) = \delta(x-x') \delta(t)
\label{eqdiffcoopx}\end{equation}
from which we obtain
 the average $\langle \cdots \rangle_{{\cal
C}}$ on diffusive trajectories
 \begin{equation} \left\langle e^{-i{2 \pi \over \phi_0} \int_{{\cal C}}
\bs{A} d {\bs l}} \right\rangle_{{\cal C}}={ \int P(x,x',t) dt
\over \int P_0(x,x',t) dt} \label{averageonC}\end{equation} 
where
 $P_0$ is solution of eq. (\ref{eqdiffcoopx}) with $B=0$, and $x$, $x'$ are taken at the extremities
 of the junction. Then the
 critical current is obtained from eq. (\ref{Icgene}).

We solve this equation with the magnetic field as a perturbation.
The eigenvalues of this diffusion equation  are solutions of
 \begin{equation}
-D \left(
\partial_x + i {2  e B y  \over \hbar} \right)^2 \psi_{n_x}= E_{n_x}  \psi_{n_x}  \ \ .\end{equation}
 and  are given by
 \begin{eqnarray} E_{n_x}&=& D Q_{n_x}^2+ D \langle \psi_{n_x}|{4 e^2 B^2 y^2
\over \hbar^2}|\psi_{n_x}\rangle  \nonumber \\
&=& D Q_{n_x}^2+ D {e^2 B^2 w^2 \over 3 \hbar^2}\ \ .
\end{eqnarray}
The new magnetic field dependent term implies an exponential decay
of the probability $P(x,x',t)=P_0(x,x',t)e^{- t / \tau_B}$, with
the characteristic time  $\tau_B$ given by \cite{AltAronov}:

\be {1 \over \tau_B}= {\pi^2 D w^2 B^2 \over 3 \phi_0^2} \
 . \label{tauB}\ee
The average (\ref{averageonC}) over  diffusive
trajectories
 is thus related to the  Laplace
transform $P_\gamma(\r,\r') =\int P_0(\r,\r',t) e^{- \gamma t} d
t$ of the probability to diffuse from one end to the sample to the
other. The numerator is  the solution of the differential equation:
\begin{equation} \left(\gamma + D {\partial \over \partial x^2} \right)
P_\gamma(x,x')  = \delta(x-x')
\label{equadifflaplace}\end{equation}
with $\gamma=1/\tau_B$,  and
with the appropriate boundary conditions. Here we assume that the
disordered junction is connected to reservoirs and  express that
the probability vanishes at the edge of the diffusive metal
(different boundary conditions could be discussed, but lead also
to the same $1/\sinh L/L_B$ of eq. (\ref{r1})). The solution of
this equation is \cite{Montambaux04}
\be P_\gamma(x,x')={L_B \over D} {\sinh x_m/L_B \sinh(L-x_M)/L_B
\over \sinh L/L_B} \label{soldiff}\ee 
where $x_m=\mbox{min}(x,x')$
and $x_m=\mbox{max}(x,x')$. We have introduced the characteristic
length:
$$L_B= \sqrt{D \tau_B}= {\sqrt{3} \over \pi} {\phi_0 \over B w} \   . $$
The coordinates $x$ and $x'$ are close to the end of the diffusive
junction. Their value is respectively $l$ and $L-l$ where the
length $l$ is of the order of the elastic mean free path $l_e$
\cite{AM2}. Since $l \ll L$, we obtain

\be P_\gamma(l,L-l)={l^2 \over D L_B \sinh L/L_B} \ .  \label{PLB}
\ee
In the limit $L_B \rightarrow \infty$, the probability scales as
$1/L$ which expresses Ohm's law that the transmission coefficient
of a diffusive system scales like the inverse of its length. Now,
from (\ref{PLB}) and (\ref{averageonC}), we obtain finally a
result independent of $l$:
\begin{equation} I_c=I_0 {L/L_B \over \sinh L/L_B}\label{r1} \end{equation}
that we write in the
final form (\ref{r2})  where $\phi= B w L$ is the flux through
the junction.
\begin{figure}[!ht]
 {\centerline{\epsfxsize=8cm
\epsffile{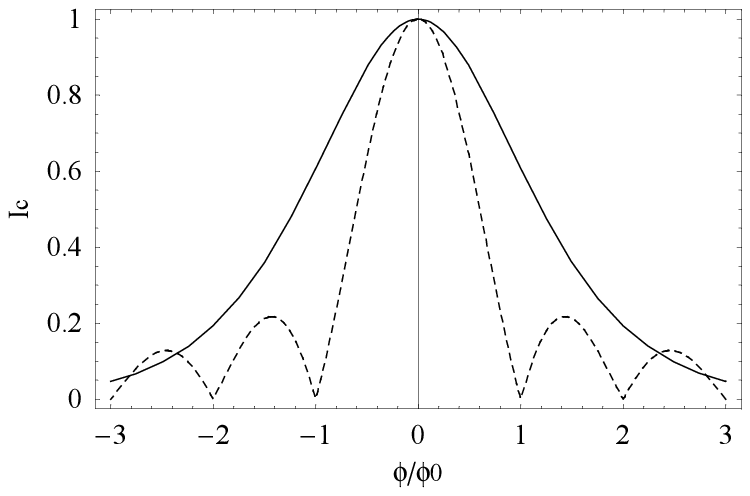}}} \caption{Comparison between the
 Fraunhofer pattern of a short ballistic junction (dotted line, eq. \ref{sinc}) and the pattern of a long ($L \gg w$)
 diffusive junction (continuous line, eq. \ref{r2}), for the magnetic field dependence
 of the critical Josephson current. $\phi$ is the flux through the junction.}\label{pattern}
\end{figure}
\medskip

{\it Phase coherence -} This result assumes full phase coherence
in the metallic junction. We now introduce a finite coherence time
$\tau_\phi$ and the probability $P_\gamma(x,x')$ is now solution
of eq. (\ref{equadifflaplace}) with $\gamma =1/\tau_B + 1
/\tau_\phi$. We immediately obtain
\be I_c = I_0 {L/L_\gamma \over \sinh L/L_\gamma} \label{Icgamma2}
\ee 
where $1/L_\gamma^2= 1/L_B^2+1/L_\phi^2$ and $L_\phi=\sqrt{D
\tau_\phi}$ is the phase coherence length. Similarly, the effect of a finite temperature can be taken into account by the thermal length $L_T=\sqrt{D/T}$, so that $1/L_B^2$ is replaced by $1/L_B^2 + 1/L_T^2$.

\medskip

{\it Gaussian accumulation of the phase - }  We now try to give a
simple interpretation of our result. The dephasing accumulated
along diffusive trajectories is characterized by the average
$\langle e^{- i \varphi} \rangle_{\cal C}$ of the phase factor
$\varphi={2 \pi \over \phi_0} \int {\bs A} \cdot d {\bs l}$ along
all diffusive paths ${\cal C}$ in the junction. Since diffusion is
a Gaussian process, the average over trajectories of a given
length, that is of a given diffusion time $t$, is
$$\langle e^{- i \varphi_t} \rangle_{\cal C} = e^{- {1 \over 2}\langle
\varphi_t^2\rangle_{\cal C}} \ . $$
 For a quasi-one-dimensional
diffusion, the average $\langle \varphi_t^2\rangle_{\cal C}$ is
simply given by $\langle \varphi_t^2\rangle_{\cal C}={4 \pi^2
\over \phi_0^2} {\overline{A^2}} \langle x_t^2 \rangle_{\cal C}$,
where $\overline{A^2}$ is an average taken along the transverse
direction. Since $\overline{A^2}= B^2 w^2/12$ and $\langle x_t^2
\rangle_{\cal C}= 2 D t$, we immediately obtain that the phase
factor averaged along all trajectories of time $t$ is
$\langle e^{- i \varphi_t} \rangle_{\cal C}= e^{- t /\tau_B}$ where
$\tau_B$ has been defined in (\ref{tauB}).  Then the dephasing has
to be averaged over all times $t$ for trajectories crossing the
sample ($x \sim l, x' \sim L-l$)

\be \langle e^{- i \varphi}\rangle_{\cal C} = {\int \langle e^{- i
\varphi_t}\rangle_{\cal C} P_0(x,x',t) dt \over \int P_0(x,x',t)
dt} \label{dephasing} \ee which is nothing but eq.
(\ref{averageonC}).
\medskip

{\it Relation with weak localization - } The magnetic field
dependence of the Josephson current probes the
 phase accumulated along diffusive trajectoires which {\it cross}
 the
sample. This physics bears of course some similarity with the weak
localization correction which probes the distribution of dephasing
along {\it closed} trajectories. Instead of probing the
probability to cross the sample $P_\gamma(l,L-l)$, the weak
localization correction probes the return probability
$P_\gamma(x,x)$. As a result, for large $L \gg L_B$, the  weak
localization correction to the dimensionless conductance (in units
of $2 e^2/h$) decays as $1/L$.
\be \Delta g= -{2  D \over  L} \int_0^L P_\gamma(x,x) dx
 \limite{L \gg L_B} -  {L_B \over L} \ ,
\ee
while in the same limit $L \gg L_B$, the Josephson current decays
exponentially
$$I_c \propto e^{-L/L_B} $$
This behavior is very reminiscent of the structure of the
harmonics of the weak localization correction on a ring as $e^{- m
L/L_B}$ (the so-called Alsthuler-Aronov-Spivak oscillations
\cite{AAS}). It is a signature of the Gaussian decay of the
probability to diffuse from one end to another after a time $t$,
which scales as $e^{-L^2/4Dt}$. That is why the weak localization
correction ($0^{th}$-harmonics) is a power law while the $m\neq 0$
harmonics decay exponentially. In the case of the ring, the
boundary conditions are periodic along the ring, leading to the
$e^{-mL/L_B}$ decay of the harmonics. Here, the trajectories can
diffuse $m$ times back-and-forth before leaving the sample,
leading to
 contributions of the form $e^{-(2m+1)L/L_B}$. The  $1/\sinh L/L_B$ behavior results obviously from the additive contributions of these diffusive trajectories :
$1/\sinh L/L_B=\sum_{m} e^{-(2 m+1) L /L_B}$.

\medskip

{\it Relation with experiments in optics - Diffusing Wave
Spectroscopy - } The Josephson relation (\ref{JJ}) involves the
transmission coefficient of  Cooper pairs which carry  random
phase factors that have to be averaged over diffusive
trajectories. It is interesting to notice a similarity between our
result for the diffusive Josephson junction  and some results obtained in
the physics of speckle correlations in optics. In optics, the
so-called Diffusing Wave Spectroscopy (DWS) is a technique
consisting in measuring the correlation function of the
transmission amplitude $t$ of light through a turbid medium,
measured at different times $0$ and $T$ \cite{DWS}. If the
scatterers of the diffusive medium can move, the correlation
function gives some information on the motion on the dynamics of
the scatterers.

 The product $\langle t(0) t^*(T) \rangle $ involves
pairings of diffusive trajectories which carry slightly different
phases, since the scatterers have moved. Therefore it measures an
average phase factor:
$$\langle t(0) t^*(T)
\rangle = \langle |t(0)|^2 \rangle \ \langle e^{-i
\varphi}\rangle_{\cal C} \ . $$
The phase accumulated along diffusive trajectories depends on the
dynamics of the scatterers. For example, for a Brownian motion of
the scatterers, the phase factor accumulated along trajectories of
time $t$ decays exponentially, $\langle e^{- i \varphi_t}
\rangle_{\cal C} = e^{- t /\tau_\gamma}$, where the characteristic
dephasing time $\tau_\gamma$ depends on the ratio between  the
wave length $\lambda$ of the incident light beam and the typical
displacement $\sqrt{D_b T}$ of the scatterers after time $T$. It
has the form $\tau_\gamma \simeq \tau_e \lambda^2 /D_b T$, where
$\tau_e$ its elastic  mean path and $D_b$ is the diffusion
coefficient for the Brownian motion of the scatterers.
Consequently the phase factor averaged over all trajectories which
cross the sample is obtained from (\ref{dephasing}) and has the
same decay (\ref{Icgamma2}) as found here for the diffusive
Josephson junction in a field. Angular and frequency speckle
correlations (the so-called $C_1$ correlations)   exhibit a similar  behavior \cite{Montambaux04,berkovits}. The common physical origin is the
Gaussian accumulation of the phase along diffusive trajectories.
\medskip

{\it Conclusion - comparison with experiments - } Surprisingly,
experiments on diffusive long diffusive $SNS$ junctions are pretty
recent, the difficulty being of keeping phase coherence along the
junction \cite{dubos}. A recent experiment have measured the
interference between two metallic $Au$ long junctions sandwiched
in a superconducting $Al$ circuit \cite{Chiodi07}. The total
current oscillates with the flux through the circuit which
modulates the relative phase between the junctions. This
interference pattern is modulated by the interference pattern of
each junction. This modulation is well described by our result.
The low field behavior is very well fitted  without
adjustable parameter by the expansion $I_c=I_0(1 - {\pi^2 \over
18} {\phi^2 \over \phi_0^2})+\cdots$ of our result (\ref{r2}). At
large field however, the decay seems to be faster than
exponential.

During the completion of this work, we have been aware of a
preprint by Hammer {\it et al.} who consider the critical current
of a long diffusive junction, within the Usadel formalism. For the
case of a perfect transmission at the NS interface, they find for
large $L/L_B$ an exponential decay of the form $L/L_B e^{-L/L_B}$
which is compatible with our equation (\ref{r1}) but they have
considered only numerically the  full range of magnetic field
\cite{Cuevas07}. Another paper \cite{Barzykin} solves Usadel equation in the limit $L \gg L_T$.

\medskip

{\it Acknowlegments - } The author acknowledges useful discussions
with L. Angers, M. Aprili, H. Bouchiat, F. Chiodi, J.C. Cuevas, S.
Gu\'eron    and M. Ferrier.

\end{document}